\documentclass[sigconf, 10pt]{acmart}

\usepackage{pifont}
\usepackage{tabularx}
\usepackage{color}
\usepackage{algorithm}
\usepackage{algpseudocode}
\usepackage{placeins}
\usepackage{subcaption}

\usepackage{alphabeta}  

\usepackage{hyperref}
 \hypersetup{
     colorlinks=true,
     linkcolor=blue,
     citecolor=magenta,
     filecolor=magenta,
     urlcolor=cyan,
     pdftitle={SCRamble}
     }
\definecolor{darkgreen}{rgb}{0.1, 0.6, 0.1}

\usepackage{ifthen}
\usepackage{calc}
\usepackage{color}
\usepackage{tikz}
\usepackage{amsfonts}
\usepackage{pgfplots}

\DeclareRobustCommand{\change}[4]{{\color{#3}#1}\ifx\relax#4\relax\else\xspace\textcolor{Gray}{\sout{#4}}\fi}

\newcommand{\figref}[1]{Figure~\ref{#1}}

\newcommand{\secref}[1]{Section~\ref{#1}}

\copyrightyear{2026}
\acmYear{2026}
\setcopyright{cc}
\setcctype{by}
\acmConference[ICDCN 2026]{27th International Conference on Distributed Computing and Networking}{January 06--09, 2026}{Nara, Japan}
\acmBooktitle{27th International Conference on Distributed Computing and Networking (ICDCN 2026), January 06--09, 2026, Nara, Japan}
\acmPrice{}
\acmDOI{10.1145/3772290.3772314}
\acmISBN{979-8-4007-1888-5/26/01}

\begin{document}
\title{SCRamble: Adaptive Decentralized Overlay Construction for Blockchain Networks}

\author{Evangelos Kolyvas}
\email{ekolyvas@aueb.gr}
\affiliation{
    \institution{Department of Informatics}
    \institution{Athens University of Economics and Business}
    \city{Athens}
    \country{Greece}
}

\author{Alexandros Antonov}
\email{aantonov@aueb.gr}
\affiliation{
    \institution{Department of Informatics}
    \institution{Athens University of Economics and Business}
    \city{Athens}
    \country{Greece}
}

\author{Spyros Voulgaris}
\email{voulgaris@aueb.gr}
\affiliation{
    \institution{Department of Informatics}
    \institution{Athens University of Economics and Business}
    \city{Athens}
    \country{Greece}
}

\renewcommand{\shortauthors}{E. Kolyvas, A. Antonov and S. Voulgaris}



\begin{abstract}

Despite being under development for over 15 years, transaction throughput
	remains one of the key challenges confronting blockchains, which typically
	has a cap of a limited number of transactions per second.
A fundamental factor limiting this metric is the network latency associated with
	the block propagation throughout of the underlying peer-to-peer network,
	typically formed through random connections.
Accelerating the dissemination of blocks not only improves transaction rates,
	but also enhances system security by reducing the probability of forks.
This paper introduces SCRamble: a decentralized protocol that significantly
	reduces block dissemination time in blockchain networks.
SCRamble's effectiveness is attributed to its innovative link selection
	strategy, which integrates two heuristics:
a scoring mechanism that assesses \emph{block arrival times} from neighboring
	peers, and a second heuristic that takes \emph{network latency} into
	account.

\end{abstract}

\maketitle
\section{Blockchain Dissemination Background}
\label{sec:background}

Blockchains are systems used to maintain a Byzantine fault tolerant state
machine, by replicating it across nodes in a Peer-to-Peer (P2P) network.

At the heart of a blockchain protocol lies the practice of grouping valid
transactions into \emph{blocks} and sharing them across the network at regular
intervals.

The dissemination network of a blockchain can be represented as an undirected
graph $G(V,E)$, with $V$ denoting the collection of \emph{nodes} and $E$
representing the set of bidirectional \emph{links} connecting the nodes.
Two nodes are referred to as \emph{neighbors} if they are connected by a link.

Blockchains usually implement a two-step block forwarding scheme where the
receiving node decides whether to request a full block.
Upon receiving a new block, a node locally validates all its transactions, and
then forwards a single packet to its neighbors containing the block header (and
as much of the block body as can fit).
This will cost 0.5 RTT (round-trip time).
By obtaining the block header within the first packet, the receiving node is
able to locally verify the header's validity before requesting the rest of the
block, i.e., check that the difficulty requirement is fulfilled in Proof-of-Work
(PoW) or that the validator is indeed deemed slot leader in Proof-of-Stake
(PoS), thereby reducing the impact of DoS attacks that aim to disseminate
invalid blocks.
In case there are additional data related to the block’s body that cannot be
transferred with the first packet, the receiving node will need to request them
from the sending node through successive pull requests.
Each pull request will cost an additional 1 RTT.
Upon receiving the full block body, the node conducts a thorough validity check
of all transactions within the block.
Block delivery is deemed complete once the node has received and thoroughly
validated the body.
If the check is successful, the node is prepared to begin forwarding the block
to its neighboring nodes (excluding the original sender) in a similar manner.
Most usually, 1.5 RTTs will be sufficient for the receiving node to get the full
block.

The dissemination of blocks in blockchains is often done through unstructured
overlay networks, which are created through random connections:
each node connects to a set of peers randomly.
The Bitcoin~\cite{bitcoin}, Ethereum~\cite{ethereum}, and Monero networks are
common examples of this type of network~\cite{propagationInBitcoin,
networkAspects, moneroNetwork}.
While this is the simplest and the easiest to implement approach, it is far from
being efficient.
Firstly, it does not evaluate existing neighbors based on their performance of
providing blocks in the recent past, allowing any lazy or poorly-connected nodes
to be part of the neighborhood as long as they wish to.
Secondly, it does not consider the proximity of neighboring nodes in terms of
network delay, potentially leading to blocks being delivered to nodes in the
same datacenter through unnecessarily long paths spanning the globe.

Minimizing the delay in message propagation can result into increased
transaction throughput by allowing for larger block sizes, a higher rate of
block generation, or the use of faster consensus algorithms.
Additionally, reducing propagation delay enhances system security by reducing
the likelihood of \emph{forks}.
A fork is the situation where two blocks are generated simultaneously, causing a
temporary uncertainty over the official chain state.
By reducing these uncertainties, which can be exploited for malicious purposes,
optimizing message propagation delay does not only boost performance but also
reinforces security measures.

\section{The SCRamble Protocol}
\label{sec:scramble}

We introduce SCRamble, an adaptive decentralized overlay construction protocol
that operates in the following manner.
Each node starts off with an arbitrary set of neighbors of size $S+C+R$,
initially picked uniformly at random, through the underlying peer sampling
service~\cite{peerSampling}.
The node splits the neighbors into two equal-sized disjoint sets:
the \emph{scoring set} (size: $S+R/2$), and the \emph{close set} (size:
$C+R/2$), and then it applies to each set the corresponding heuristic.

\subsection{The Scoring heuristic}
\label{sec:score_heuristic}

A node builds its \emph{scoring set} by applying the scoring heuristic:
It observes how fast each neighbor relays blocks to it, and it assigns a score
to each node in this set based on its performance.
Specifically, for each new block that the node receives, if $t_1$ is the time it
receives the block for the first time from a peer $u$ ($u \in$ scoring set), and
$t_2$ is the time it receives the same block for the second time from another
peer $v$ ($v \in$ scoring set, $u \neq v$), the node will add the difference
between these two timestamps in milliseconds ($t_2 - t_1$) to the score of the
first peer $u$ as points.
Both $t_1$ and $t_2$ refer to the local time of the receiving node.
Except for the peer who managed to deliver the new block first, all the other
peers of the scoring set will receive no points from this block.
Periodically (every $k$ blocks), the node will rank the nodes of the scoring set
by their cumulative score (from the past $k$ blocks).
It will retain the top $S$ nodes with the highest scores and disconnect from the
$R/2$ nodes with the lowest scores.
Additionally, it will make connections to $R/2$ random new neighbors as a way of
refreshing the scoring set with new unseen neighbors, potentially discovering
better-performing nodes in the rest of the network.
The scores of all nodes in the updated scoring set will be reset to zero,
including those nodes retained from the previous round.
This process is repeated every $k$ blocks.
Algorithm~\ref{alg:algo_s} shows the pseudocode of our scoring heuristic.

\begin{algorithm}
\caption{Scoring heuristic}
\label{alg:algo_s}

\begin{algorithmic}[1]

\State \textit{// PSS = the underlying Peer Sampling Service}
\State \textit{// responsive = alive and able to handle another connection}
\State
\State \textit{// Rejuvenate node's Scoring Set (SS)}
\State \textit{// CS = node's Close Set}
\State
\State \textit{// $u, v \in$ SS, $u \neq v$}
\State \textit{// $t_1$ = 1st time of receiving a new block from peer $u$}
\State \textit{// $t_2$ = 2nd time of receiving the same block from peer $v$}
\State \textit{// dt = ($t_2 - t_1$) in msec}
\State

\Loop{} periodically
    \For{$k$ blocks}
        \State add $dt$ points to the score of peer $u$
    \EndFor

    \State

    \While{$SS.size > S$}
        \State remove the lowest point node from $SS$
    \EndWhile

    \State

    \While{$SS.size < (S+R/2)$}
        \State pick a random node $w$ from $PSS$
		\If {$w$ is responsive $AND$ $w \notin$ ($SS \cup CS$)}
            \State add $w$ to the $SS$
        \EndIf
    \EndWhile

    \State

    \State set all points to zero
\EndLoop

\end{algorithmic}
\end{algorithm}

In contrast to manually designed protocols that often necessitate extensive
tuning of parameters for individual blockchain networks, we design a scoring
function that is directly matched to the objective that we are trying to
optimize and automatically identifies the best topology for any network setting.
The scoring heuristic automatically adapts not only to network heterogeneity,
such as latencies, bandwidth, congestion between nodes, and nodes’ generic
computational abilities, but also it captures consensus-layer characteristics,
such as mining power or good connectivity with mining pools.

\subsection{The Close heuristic}
\label{sec:close_heuristic}

For the \emph{close set}, SCRamble applies the close heuristic:
Our second heuristic operates one layer beneath our first heuristic, i.e., at
the networking/peer-to-peer layer.
Thus, it is unaware of any consensus layer characteristics, such as who is
generating or transmitting new blocks faster.
The node measures its latency to any other peer in the close set using a number
of ping messages.
Periodically (every few seconds), it will average the latencies to each peer in
the close set, retain the $C$ nodes with the lowest average latency, and discard
the $R/2$ nodes with the highest average latency, replacing them with $R/2$
randomly picked ones.
Similarly to the logic of the scoring heuristic, this is a way of refreshing the
close set with new unseen neighbors, potentially discovering closer neighbors
than the current ones.
All current latencies will be forgotten, including those to nodes retained from
the previous round.
This process is going to repeat every few seconds.
The periodic rejuvenation of a node’s neighbor set helps it adapt to dynamic
conditions.
Naturally, this heuristic brings us nodes in proximal geographical locations,
although our heuristic does not take location into account and focuses solely on
network latency.
Algorithm~\ref{alg:algo_c} shows the pseudocode of our close heuristic.

\begin{algorithm}[t]
\caption{Close heuristic}
\label{alg:algo_c}

\begin{algorithmic}[1]

\State \textit{// PSS = the underlying Peer Sampling Service}
\State \textit{// responsive = alive and able to handle another connection}
\State
\State \textit{// Rejuvenate node's Close Set (CS)}
\State \textit{// SS = node's Scoring Set}
\State

\Loop{} periodically
    \ForAll{node $v \in CS$}
        \State measure the RTT to node $v$
    \EndFor

    \State

    \While{$CS.size > C$}
        \State remove node with highest avg RTT from $CS$
    \EndWhile

    \State

    \While{$CS.size < (C+R/2)$}
        \State pick a random node $w$ from $PSS$
        \If {$w$ is responsive $AND$ $w \notin$ ($SS \cup CS$)}
            \State add $w$ to the $CS$
        \EndIf
    \EndWhile

    \State

    \State set all latencies to zero
\EndLoop

\end{algorithmic}
\end{algorithm}

In general, rapid dissemination at a global scale has to deal with the following
two challenges:
First, a new block should be distributed fast and exhaustively at a local scope,
harnessing the low-latency links of geographically proximal locations and
ensuring that every single nearby node receives the block through a fast,
low-latency, local path.
Second, a dissemination algorithm should incorporate a global perspective,
ensuring rapid distribution of the block to remote areas.

Intuitively, a block generated in Tokyo should quickly reach a node in Osaka
through local connections, instead of taking a longer route that goes through
Dublin.
Meanwhile, the node in Dublin should receive the block fairly quickly via a
direct shortcut from the vicinity of Tokyo to a location in Western Europe,
instead of having to wait for the block to gradually traverse all of Asia and
Europe through numerous local dissemination steps.

\begin{figure}[t]
  \includegraphics[width=\columnwidth]{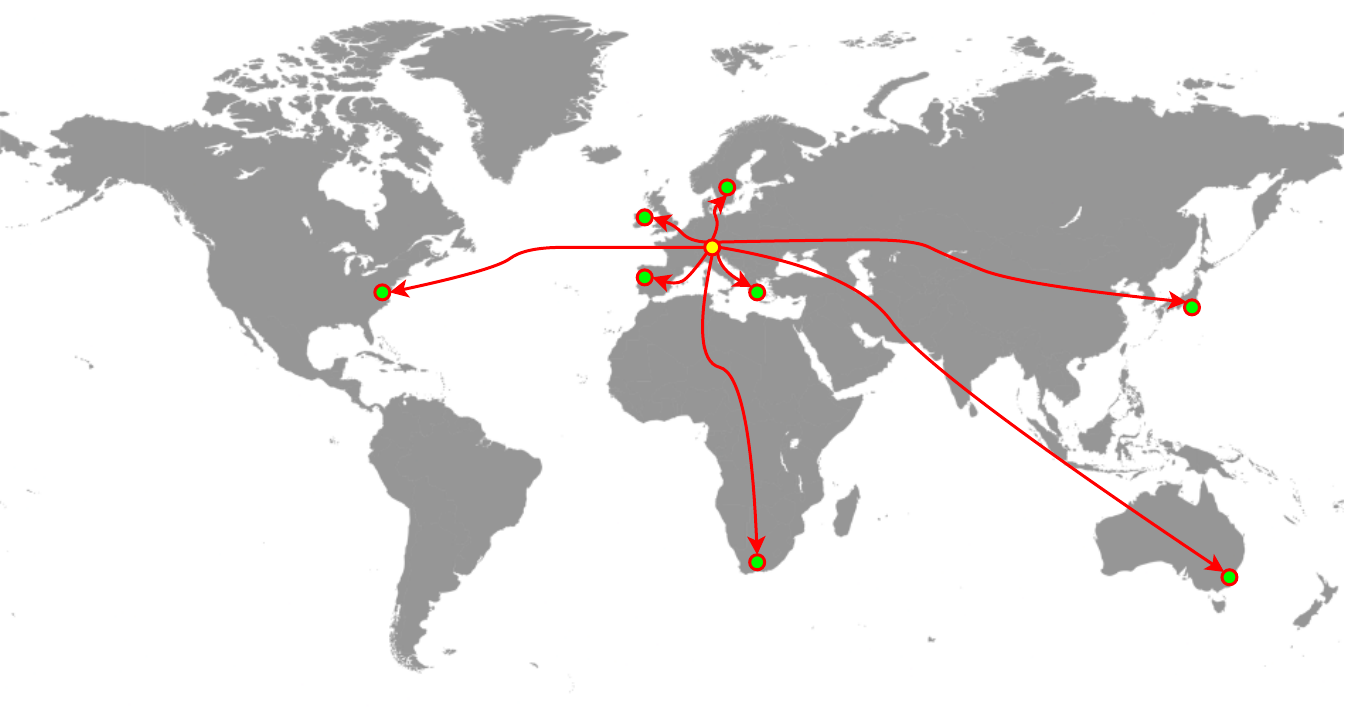}
  \caption{Blocks should be propagated to a few nearby and to a few distant
	nodes (close heuristic).}
  \label{fig:world}
\end{figure}

\figref{fig:world} illustrates an overview of the suggested approach behind our
close heuristic.
Blocks should be transmitted over both local, low-latency connections, and
remote, long-distance ones.\newline

In summary, each node keeps three peer sets that work together to optimize block
dissemination:
\begin{itemize}
\item \textbf{Scoring neighbors} are those that have consistently relayed new
	blocks the fastest, ensuring reliable data exchange.
\item \textbf{Close neighbors} are chosen for having the lowest measured network
	latency, facilitating rapid, low-latency communication within close
		proximity.
\item \textbf{Random neighbors} are picked uniformly at random to introduce
	global shortcuts, and to discover new peers.
\end{itemize}

By combining high-performance links, low-latency connections, and random
shortcuts, SCRamble achieves fast block propagation.

\section{Experimental Setup}
\label{sec:experimental_setup}

In order to emulate realistic network latencies between nodes, we acquired a
real-world latency trace publicly available by
WonderNetwork~\cite{wondernetwork}, and we turned it into a dataset reflecting
the geographic distribution of Bitcoin network nodes~\cite{bitnodes}, adopting
the way followed in CougaR~\cite{cougar}.
All evaluations were conducted using the Peer-Net Simulator~\cite{peernet}.

In all evaluation, the notation S$x$-C$y$-R$z$ represents a configuration where
each node establishes $x$ scoring, $y$ close, and $z$ random links.

\begin{figure}[t]
  \includegraphics[width=\columnwidth]{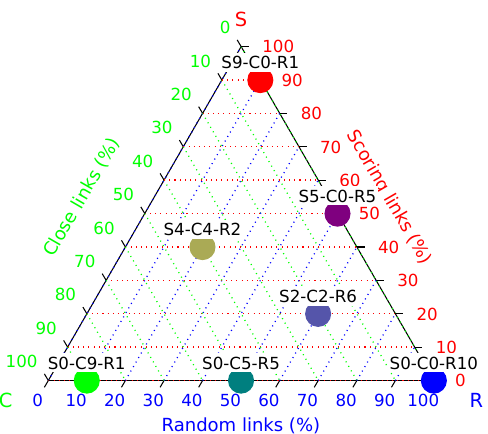}
  \caption{Links are shared among three different sets:
	Scoring (S), Close (C), and Random (R). Experimental space with 10 links.}
  \label{fig:triangle}
\end{figure}

\figref{fig:triangle} portrays the parameter space under investigation.
The blue vertex represents the configuration where only random links are being
established, i.e., none of the two heuristics is being applied.
The red and green vertices represent the configurations where only scoring-based
or only close-based links are being established, respectively.
In these last two scenarios, we still include one randomly selected link, which
is necessary for refreshing the links in both heuristics.
The edge between the red and blue vertices represents setups with only the
scoring heuristic, while the edge between the green and blue vertices represents
setups with only the close heuristic.
Moving along the edges towards a certain vertex, the ratio of nodes in favor of
that vertex increases at the expense of the vertex we are leaving behind.
All points in the triangle's interior correspond to setups that combine both
heuristics.

Note that the color coding in \figref{fig:S_vs_C_vs_R} matches the respective
parameter space coloring of \figref{fig:triangle}.

\section{Evaluation of SCRamble's Link Selection Policy: Scoring vs. Close vs. Random}
\label{sec:evaluation}

\begin{figure*}[h!]
    \hbox{\begin{subfigure}{0.33\linewidth}
        \includegraphics{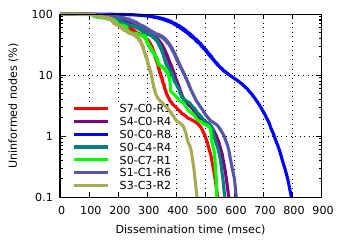}
        \caption{}
        \label{fig:050msec_15RTT}
    \end{subfigure}%
    \begin{subfigure}{0.33\linewidth}
        \includegraphics{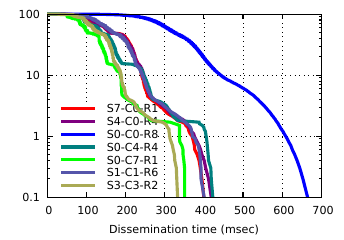}
        \caption{}
        \label{fig:020msec_15RTT}
    \end{subfigure}%
    \begin{subfigure}{0.33\linewidth}
        \includegraphics{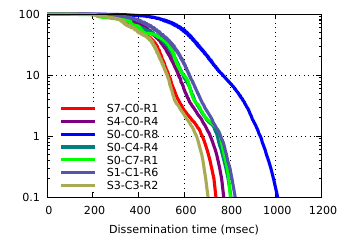}
        \caption{}
        \label{fig:100msec_15RTT}
    \end{subfigure}}%
	\begin{subfigure}{0.33\linewidth}
        \includegraphics{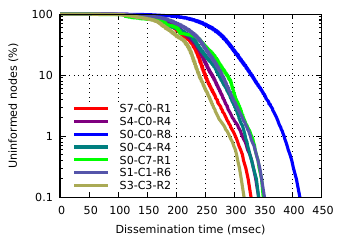}
        \caption{}
        \label{fig:050msec_05RTT}
    \end{subfigure}%
    \begin{subfigure}{0.33\linewidth}
        \includegraphics{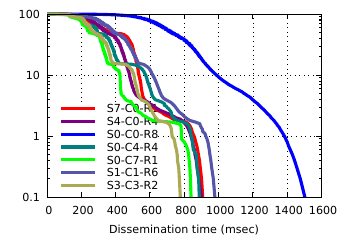}
        \caption{}
        \label{fig:050msec_35RTT}
    \end{subfigure}%

	\caption{Dissemination progress over time for different setups of body
	validation delay and total RTTs per block transfer respectively.
	(a) 50 msec, 1.5 RTTs (default scenario), (b) 20 msec, 1.5 RTTs (lower
	delay), (c) 100 msec, 1.5 RTTs (higher delay), (d) 50 msec, 0.5 RTTs (fewer
	RTTs), (e) 50 msec, 3.5 RTTs (more RTTs)}
    \label{fig:S_vs_C_vs_R}
\end{figure*}

In this section, we evaluate SCRamble by examining its performance across
alternative options for the link selection policy, specifically the percentages
of scoring, close, and random neighbors within the neighbor set.
We ran multiple experiments to assess various combinations of scoring, close,
and random links across different setups involving block validation delays and
block sizes.

\figref{fig:S_vs_C_vs_R} illustrates
the dissemination progress over time since a block has been generated, showing
the percentage of nodes that have not yet received and validated the block.
Each line in the plots corresponds to a distinct experiment, averaging the
dissemination of 100 blocks from uniformly randomly selected miners.
The experiments were conducted with scoring links included, are shown after 128
rounds of calibration, with 100 blocks per round, to achieve long-term
performance convergence.
Line colors represent setups with the percentages of scoring, close, and random
neighbors, corresponding to the respective colors in \figref{fig:triangle}.

\figref{fig:050msec_15RTT} shows a setting with the default delay
values\footnote{\url{https://statoshi.info/d/000000003/function-timings}}, i.e.,
5\,msec and 50\,msec per node for the header and body validation delays,
respectively, and 1.5 RTTs for a block transfer (as mentioned in
\secref{sec:background}).

\figref{fig:020msec_15RTT} and \figref{fig:100msec_15RTT} consider two
representative sets of experiments with different body validation delays:
one set with a lower delay (20 msec) than the usual (50 msec), and another one
with a higher delay (100 msec).

To assess the effect of the extra RTTs incurred by larger blocks on
dissemination, as well as the scenario in which the block body fits within the
initial packet along with the header, allowing the entire block to be
transferred in 0.5 RTTs, we conducted experiments with a wide range of total
RTTs required per block transfer.
\figref{fig:050msec_05RTT} and \figref{fig:050msec_35RTT} report two
representative sets of experiments:
one set with fewer total RTTs (0.5 RTTs), and another one with more (3.5 RTTs),
compared to the usual (1.5 RTTs).

We observe that when neither of the two heuristics is applied, exclusive use of
random links (S0-C0-R8) yields the worst performance.
Using just one of our heuristics and applying it to half of the neighbors
(S4-C0-R4, S0-C4-R4), or using both heuristics and applying them to a minority
of the neighbors (S1-C1-R6), provides better results than 100\% random, however,
we can do better.
The reason is that in all three of the above cases, the percentage of random
nodes remains high (i.e., R $\ge50$\% of the neighbors).
This is proven by the fact that using just one of our heuristics and applying it
to the majority of the neighbors (S7-C0-R1, S0-C7-R1) provides even better
results.
When the node delay increases, or the total number of RTTs decreases, scoring
peers yield better results.
Vice versa, when the node delay decreases, or the total number of RTTs
increases, close peers have an advantage.
However, note that setups involving only close nodes are very likely to split
the network into disconnected components due to the nodes' greedy policy to team
up exclusively with nearby nodes.
The best results are achieved by applying both heuristics to the majority of
each of the scoring and close sets (S3-C3-R2).

\section{Conclusions}
\label{sec:conclusions}

We introduced SCRamble:
an adaptive decentralized overlay construction protocol that determines which
neighbors a node should connect to in order to minimize block dissemination time
in blockchain networks.
SCRamble's link selection policy consists of three key elements:
\emph{(a) neighbors' recent record of providing new blocks}, captured by a
sophisticated scoring function,
\emph{(b) proximity}, in terms of network latency, and
\emph{(c) randomness}.
We emphasized the significance of incorporating each of these components into
the link selection policy up to a certain threshold, ensuring they contribute
their beneficial characteristics without exceeding that limit, as doing so would
lead to unnecessary link usage without any benefits.
In this context, we assessed the effects of varied block validation delays and
sizes.

\begin{acks}

Work funded by \emph{Input Output Global} (\emph{IOG}) in the context of the
\emph{``Eclipse-Resistant Network Overlays for Fast Data Dissemination''}
project, to optimize Cardano's overlay network.

\end{acks}

\bibliographystyle{ACM-Reference-Format}
\bibliography{./refs}

@inproceedings{bitcoin,
	title={Bitcoin: A peer-to-peer electronic cash system},
	author={Nakamoto, S.},
	year={2008}
}

@inproceedings{ethereum,
	title={A next-generation smart contract and decentralized application platform},
	author={Buterin, V. and others},
	journal={white paper},
	year={2014}
}

@inproceedings{propagationInBitcoin,
	title={Information propagation in the bitcoin network},
	author={Decker, C. and Wattenhofer, R.},
	booktitle={IEEE P2P 2013},
	pages={1--10},
	year={2013}
}

@inproceedings{networkAspects,
  title={Network layer aspects of permissionless blockchains},
  author={Neudecker, T. and Hartenstein, H.},
  journal={IEEE Communications Surveys \& Tutorials},
  year={2018}
}

@inproceedings{moneroNetwork,
  title={Exploring the monero peer-to-peer network},
  author={Cao, T. and Yu, J. and Decouchant, J. and Luo, X. and Verissimo, P.},
  booktitle={International Conference on Financial Cryptography and Data Security},
  pages={578--594},
  year={2020},
  organization={Springer}
}

@inproceedings{peerSampling,
	title={Gossip-based peer sampling},
	author={Jelasity, M. and Voulgaris, S. and Guerraoui, R. and Kermarrec, A-M and Van Steen, M.},
	journal={ACM TOCS},
	year={2007}
}

@online{wondernetwork,
	title = {{A day in the life of the Internet (WonderProxy)}},
	author = {{Paul Reinheimer}},
	note = {\url{https://wonderproxy.com/blog/a-day-in-the-life-of-the-internet/}},
	year = {2024}
}

@online{bitnodes,
	title = {Bitnodes},
	note = {\url{https://bitnodes.io/}},
	year = {2024}
}

@inproceedings{cougar,
	author = {Kolyvas, Evangelos and Voulgaris, Spyros},
	title = {CougaR: Fast and Eclipse-Resilient Dissemination for Blockchain Networks},
	year = {2022},
	isbn = {9781450393089},
	publisher = {Association for Computing Machinery},
	url = {https://doi.org/10.1145/3524860.3539805},
	doi = {10.1145/3524860.3539805},
	booktitle = {Proceedings of the 16th ACM International Conference on Distributed and Event-Based Systems},
	pages = {7–18},
	numpages = {12},
	location = {Copenhagen, Denmark}
}

@online{peernet,
	title = {{Peer-Net Simulator}},
	note = {\url{https://github.com/PeerNet}},
	year = {2024}
}

\end{document}